\documentclass{ws-jai}
\usepackage[flushleft]{threeparttable}
\begin{document}

\catchline{}{}{}{}{} 

\markboth{Federico Nati}{POLOCALC: a Novel Method to Measure the Absolute Polarization Orientation\\
of the Cosmic Microwave Background}

\title{POLOCALC: a Novel Method to Measure the Absolute Polarization Orientation\\
of the Cosmic Microwave Background}


\author{Federico Nati$^{1}$, Mark J. Devlin$^{1}$, Martina Gerbino$^{2}$, Bradley R. Johnson$^{3}$, Brian Keating$^{4}$, Luca Pagano$^{5}$, Grant Teply$^{4}$}

\address{
$^{1}$Department of Physics and Astronomy, University of Pennsylvania, Philadelphia, 19104 PA, USA\\
$^{2}$The Oskar Klein Centre for Cosmoparticle Physics, Department of Physics, Stockholm University, SE-106 91 Stockholm, Sweden\\
$^{3}$Department of Physics, Columbia University, New York, NY 10027, USA\\
$^{4}$Department of Physics, University of California, San Diego, CA 92093-0424, USA\\
$^{5}$Institut d'Astrophysique Spatiale, CNRS, Univ. Paris-Sud, Universit\'{e} Paris-Saclay, B\^{a}t. 121, 91405 Orsay cedex, France\\
}

\maketitle

\corres{$^{1}$Corresponding author.}

\begin{history}
\received{2017 April 7};
\accepted{2017 April 27}
\end{history}

\begin{abstract}
We describe a novel method to measure the absolute orientation of the
polarization plane of the Cosmic Microwave Background (CMB) photons with
arcsecond accuracy, enabling unprecedented measurements for cosmology and
fundamental physics. Existing and planned CMB polarization instruments looking
for primordial B-mode signals need an independent, experimental method for
systematics control on the absolute polarization orientation. The lack of such
a method limits the accuracy of the detection of inflationary gravitational
waves, the constraining power on the neutrino sector through measurements of
gravitational lensing of the CMB, the possibility of detecting Cosmic
Birefringence, and the ability to measure primordial magnetic fields. Sky
signals used for calibration and direct measurements of the detector
orientation cannot provide an accuracy better than 1$^{\circ}$. Self-calibration
methods provide better accuracy, but may be affected by foreground signals and
rely heavily on model assumptions, losing constraining power on fundamental
processes, like Cosmic Birefringence, Faraday Rotation and chiral gravity
models. The POLarization Orientation CALibrator for Cosmology, POLOCALC, will
dramatically improve instrumental accuracy by means of an artificial
calibration source flying on high-altitude balloons and aerial drones.
Polarization angle calibration requires observation of a well-characterized
distant source at high elevation angles. A balloon-borne calibrator will
provide a source in the far field of larger telescopes, while an aerial drone
can be used for tests and smaller polarimeters. POLOCALC will also allow a
unique method to measure the telescopes' polarized beam. Even a two-hour
balloon flight will allow enough time to perform polarization angle calibration
and polarized beam function measurements. The source will make use of both
narrow and broad-band microwave emitters between 40 and 150 GHz coupled to
precise polarizing filters. The orientation of the source polarization plane
will be registered to absolute celestial coordinates by star cameras and
gyroscopes with arcsecond accuracy. This project can become a rung in the
calibration ladder for the field: any existing or future CMB polarization
experiment observing our novel polarization calibrator will enable measurements
of the polarization angle for each detector with respect to absolute sky
coordinates.
\end{abstract}


\section{Introduction}
\noindent
Existing and planned Cosmic Microwave Background (CMB) polarization instruments
need an independent, experimental method for exquisite systematics control on
the absolute polarization orientation. The lack of such a method limits the
accuracy on the detection of Inflationary Gravitational Waves, the constraining
power on the neutrino sector through measurements of gravitational lensing of
the CMB, the possibility to detect Cosmic Birefringence (CB) that would
represent a paradigm shift in fundamental physics, and the ability to measure
Primordial Magnetic Fields.

\subsection{State-of-the-art}
CMB intensity and polarization measurements have been an invaluable resource
for testing cosmological models and fundamental physics, since processes that
operated in the early Universe, or acted on the photons' way to the Earth, left
very weak but distinct imprints on the uniform background. We have learned that
space-time is flat, that our Universe is 13.8 billion years old and its energy
content is dominated by cold dark matter and dark energy
\cite{planck14a,planck16a}. The polarization anisotropy patterns are a
combination of even-parity (E) and odd-parity (B) modes. According to the
standard cosmological model, first order density (scalar) perturbations at the
last scattering surface produced both intensity and E-mode polarization
anisotropies, while B-modes require tensor perturbations with parity-odd
components. Recent E-modes polarization measurements are consistent with the
standard cosmological model \cite{planck14b,planck16b,louis16}.
Primordial B-mode signals have never been detected. They are fainter and can be
easily contaminated, yet they may reveal crucial information on our Universe.
Many major questions may find their answers in the subtle B-mode signals, like:
did inflation really happen? What is the signal level of primordial
gravitational waves predicted by inflation? How many neutrino species are there
and what is their mass? Were magnetic fields already present in the early
Universe? Where do magnetic fields in galaxies and galaxy clusters come from?
Accurate measurements of B-modes can also reveal (or place limits on) Cosmic
Birefringence, a revolutionary departure from the Standard Model that allows to
probe the validity of fundamental symmetries, and to investigate the nature of
Dark Energy and test extensions of the General Relativity. To move forward,
current and future CMB polarization experiments are aiming at an unprecedented
level of sensitivity, therefore systematic effects that until today were less
important than statistical uncertainties are becoming the most significant
limitation. An apparent polarization rotation produces a power leakage from
E-mode spectrum into B-mode signal that results in a systematic signal in the
B-modes spectrum and in the TB and EB cross spectra encoding the correlation of
temperature (T) and E-mode signals with B-modes. As a road map for ground
experiments and future satellite missions looking for primordial B-modes has
been outlined \cite{s416,divalentino16}, it is time to overcome existing
experimental methods to calibrate the polarization angle that only allow a
total accuracy around 1$^{\circ}$ \cite{abitbol16,planck16c,debernardis16}.
Such a systematic noise has in fact emerged above existing and forecasted
sensitivity levels, as we will show later.

\subsection{Astrophysical and Instrumental Origins of B-modes}
Origins of B-modes can be either primordial or more recent and they can be either
astronomical or instrumental. Several effects also produce non-vanishing TB and
EB cross angular cross power spectra, including not only instrumental
systematics, but also genuine physical phenomena affecting the orientation of
the polarization plane of the CMB photons and galactic foregrounds, limiting
the accuracy of self-calibration methods which rely on the assumption that the
origin of TB and EB spectra is purely instrumental \cite{keating13}. 
Spurious B-modes originated from polarization systematics convert into a bias
affecting cosmological information \cite{shimon08,odea07,mactavish08,hu03}. In
particular, a rotation of the polarization sensitive detector with respect to
the reference frame used to define the cosmological Stokes parameters generates
B-modes from E-modes. A miscalibration of the orientation of the polarimeter
detectors is equivalent to such a coherent rotation. Any cosmic or instrumental
polarization rotation converts E-mode into B-mode and vice versa, but the
E-mode signals are significantly larger than the B-modes, so the net result is
an excess on the B-mode power spectrum. An unknown bias on the polarization
angle limits the accuracy of B-mode measurements, weakening the constraints on
cosmological parameters, and it has a complete degeneracy with genuine physical
phenomena like CB \cite{yadav12,pagano09}. While the two angular coordinates
defining the CMB polarimeter's pointing direction can be calibrated with high
accuracy, the third angle defining the rotation of the polarization plane along
the detector's line of sight is much harder to refer to absolute celestial
coordinates. Also non-ideal telescope beam functions mix intensity and
polarization signals, resulting in large polarization systematics if not
properly detailed \cite{jones07,miller09,odea07}. In some
cases these kinds of irregularities can also produce TB or EB signals.  Since
the current experimental techniques are not able to significantly reduce the
level of polarization angle systematics, many experiments make use of a
self-calibration technique, based on the assumption that any TB or EB power in
the cross spectra comes from an instrumental bias \cite{keating13,kaufman14a}.
However genuine correlation between E and B-modes spectra from astrophysical
signals limit the accuracy of self-calibration \cite{abitbol16}. Besides,
self-calibration comes at the price of losing constraining power on fundamental
phenomena that can affect the polarization orientation of the CMB photons, like
Cosmic Birefringence.

One of inflation's predictions is a background of gravitational waves that
propagated in the primordial Universe. Inflationary Gravitational Waves (IGW)
were tensor fluctuations that imprinted E and B-mode polarization patterns in
the CMB. We are still looking for evidence of this key prediction. Recent
measurements place a limit on the ratio between the tensor to scalar
perturbations produced by IGWs of $r < 0.07$ \cite{planck16a, ade16}.
Simulations show that IGW searches targeting tensor-to-scalar ratios of $r
\simeq 0.01$ must calibrate the relationship between the instrument frame and
the reference frame on the sky at least with arcminute precision
\cite{shimon08,odea07,mactavish08}. Sub-arcminute accuracy will be required
for even smaller values of r, targeted by CMB-S4 and future satellite missions.

More recent processes strongly contaminate the subtle primordial
polarization imprints. Intervening large scale structures can change the
polarization properties of the CMB photons through gravitational lensing, which
also encodes information on the total neutrino mass in the CMB polarization
patterns. Precise lensing measurements place the most stringent limits on the
neutrino sector and are necessary to separate gravitational distortions from
the primordial signals 
\cite{planck14c,planck16d,planck16e,sherwin16,vanengelen16,madhavacheril15}.

Galactic dust polarized emission dominates primordial B-mode signals at all
latitudes \cite{planck16f} and represents the most pernicious
astronomical foreground that needs to be accurately subtracted to clean the
images of the early Universe. Polarized galactic foregrounds produce EE, BB, TB
and EB angular power spectra. Even small non-vanishing cross spectra limit the
accuracy of self-calibration methods \cite{abitbol16}. 

Cosmic Birefringence predicts a photon polarization plane rotation due to
non-standard-model phenomena coupling with electromagnetism. CB can break
parity symmetry via a Chern-Simons coupling term \cite{glusevic10} or violate
the Einstein Equivalence Principle via a pseudo-scalar field interaction with
light \cite{carroll90}. CB rotation converts E-modes into B-modes and produces
TB and EB cross spectra. Cosmic Birefringence detection would be a paradigm
change in fundamental physics. CMB is the most distant polarized light we can
observe, so it can accumulate even a very small CB rotation over cosmological
distances \cite{xia10,yadav12}. While it is possible to isolate CB from other
effects breaking the degeneracy through frequency or scale dependencies, its
effect is indistinguishable from a bias on the measured polarization angle,
which therefore requires an accurate independent calibration
\cite{pagano09,rosset10}. Upper limits to CB from CMB polarization data have
been measured \cite{planck16g,ade15,gruppuso16}. Since the net result of CB is to
increment the B-mode spectrum, if we do not take into account its effect we
introduce a positive bias for $r$. It is also important to notice that if CB
starts to accumulate during photon propagation right after recombination,
before gravitational lensing mixes temperature and polarization signals, then
lensed CMB spectra do not contain only primordial anisotropies, but they also
encode a rotation effect due to birefringence \cite{gubitosi15}.

If magnetic fields were present in the primordial plasma, they affected the
polarization angle of CMB photons through Faraday Rotation (FR). In other
words, FR allows to measure magnetic fields in the early Universe through the
detection of a rotation of the polarization orientation. Like other rotation
effects, FR produces E-mode to B-mode leakage \cite{kosowsky96,pogosian12} and,
similarly to CB, generates non-vanishing TB and EB spectra. While some
explanations for galactic microgauss fields are provided by dynamo mechanisms,
the origin of magnetic fields in large structures is still unclear. Accurate
CMB measurements can place bounds on Primordial Magnetic Fields (PMF) and
provide explanations on their origin. Since this is a primordial effect, it
mixes with the tensor B-modes caused by IGWs. FR can be distinguished from
other phenomena thanks to its frequency dependence.  Existing telescopes
already have the sensitivity to detect mode-coupling correlations sourced by a
scale-invariant PMF below 1 nG or even better when the weak lensing
contribution is subtracted. But an independent and improved polarization angle
calibration is required. Planck measured an upper limit for the PMFs to 4.4
nanogauss (nG) \cite{planck16h} assuming zero helicity, and a similar result
was confirmed by POLARBEAR \cite{ade15}. By assuming maximal helicity, the
Faraday Rotation only constraint on stochastic PMFs measured by Planck is
\textless1380 nG \cite{planck16h}, obtained using only large scale EE and BB
modes at 70 GHz. The current limits on PMFs are usually obtained assuming zero
or maximally helical field. Accurate measurents of TB and EB
spectra will open the possibility of constraining the helicity of PMFs
\cite{ballardini15}.

\section{Limitations of the existing calibration methods}
The limitations of the current methods for calibrating the absolute
polarization direction of the detectors are summarized in the following list.
\begin{arabiclist}[(5)]
\item Lack of natural reference sources. Sky signals from known sources are
traditionally used to calibrate the intensity of CMB experiments, but there is
no analogous standard for polarized beam measurements or polarization angle
calibration. The few astronomical candidates suffer from frequency dependence
and time variability. They are not visible from all observatories and are
extended sources. The best option is Tau-A which allows an accuracy for the
polarization orientation between 1$^{\circ}$ and 0.5$^{\circ}$
\cite{planck16i,polarbear14,weiland11}. 
\item Given their typical dimensions, the alignment of detector arrays and cold
optics with better than 1$^{\circ}$ accuracy requires a positioning uncertainty
smaller than 1mm \cite{debernardis16}. Microwave detectors must be cooled down
from 300 K to 100 mK.  Differential contractions of the materials in the cryostat
introduce additional misplacements larger than 1 mm. Even with a careful design
to mitigate these effects it is hard to fully recover the accuracy, since all
the parts must be mounted together at room temperature. During
operations external pressure and temperature can change, affecting the cryostat
internal conditions. It is also critical to refer the detector orientation with
respect to the telescope and the receiver mount once the cryostat is closed. As
a result, direct polarization angle calibration is not possible with an
accuracy better than 1$^{\circ}$.
\item Many experiments require the use of polarization modulator systems based
on large rotating half wave plates (HWP) \cite{simon16,thornton16,lspe12}.
Thanks to the large filter diameter, in principle they can calibrate the
polarization angle with good accuracy. However, ideal, wide-band, optically
uniform and thermally stable HWPs do not exist, so they introduce uncontrolled
biases degrading the accuracy \cite{pisano14, essinger16}. Some experiments make use of
reflective polarization modulation systems, but also in this case non-ideal
properties weaken the alignment accuracy \cite{miller16,chuss12,houde01}. HWP
rotating mount can be also used to place a thin film polarization grid in front
of the receiver and calibrate the absolute polarization orientation of the
detectors \cite{koopman16}. However, any strategy based on optical elements
placed between the mirrors and the polarimeter does not allow to measure the
polarized beam systematics induced by the warm optics. 
\item The characterization of the polarized beams requires a calibrated,
polarization-pure source placed in the far field of the telescope. Even the
best candidates among the astronomical sources that could be used for this
scope, like Tau-A \cite{planck16i,polarbear14,weiland11}, are not point-like
sources, and their frequency spectrum within the sensitive bands of the
polarimeters is not precisely known.
\item The standard cosmological model predicts that in the early Universe the
odd-parity and the even parity signals should be completely unrelated. The
angular power spectrum is commonly used to study the statistical properties of
the CMB anisotropies. It represents the variance of the signal fluctuations in
temperature (TT) and polarization (EE, BB) for different angular scales. The
cross spectra indicated as TE, TB and EB, measure the correlation across
temperature and polarization modes. The standard cosmological model predicts
that TB and EB identically vanish, which means that the odd-parity signal (B)
should be completely unrelated to intensity (T) and even-parity signals (E) in
the early Universe. This prediction can be used to calibrate CMB polarimeters
through a self-calibration method \cite{keating13,kaufman14a}, at the expense
of losing detection capability on genuine physical quantities. This method is
particularly effective for high resolution and/or high sky coverage
experiments. However, the initial assumption is not true in the presence of
phenomena that produce non-vanishing TB and EB. In these cases self-calibration
loses accuracy and introduces biases on cosmological parameters
\cite{abitbol16}. Besides, this method destroys the possibility to measure or
to place limits on phenomena that generate TB and EB spectra, like Cosmic
Birefringence, Faraday Rotation and chiral gravity models \cite{kaufman14b,gerbino16}. TB
and EB correlations can also be introduced by polarized galactic foregrounds
and instrumental systematics. 
\end{arabiclist}

\section{A novel method: POLOCALC}
POLOCALC will overcome current limitations by implementing a novel approach to
calibrate the polarization angle and the beam pattern of the CMB polarimeters
with an accuracy between 0.01$^{\circ}$ and 0.001$^{\circ}$. Such a method will
fully enable the constraining power of the CMB data set with no change to
existing or planned instruments.
\subsection{Objectives}
POLOCALC objectives are divided as follows:
\begin{romanlist}[(iiii)]

\item Enabling unprecedented accuracy on the measurement of Inflationary
Gravitational Waves and the tensor-to-scalar ratio through a novel calibration
method insensitive to polarized foreground emission;

\item Controlling systematics on CMB gravitational lensing signals,
improving the accuracy on gravitational potential measurements and the
resulting constraints on the neutrino sector;

\item Revealing (or constraining) Cosmic Birefringence and physics
beyond the Standard Model that can violate fundamental symmetries;

\item Measuring Primordial Magnetic Fields through Faraday Rotation,
anchoring inflationary models and providing evidence of the seeds that
originated magnetic fields in large structures.

\end{romanlist}
POLOCALC will utilize two celestial coordinates (from the accurate star camera
pointing direction) to determine the third angle defining the rotation of the
polarization plane along the detectors' line of sight. The camera pointing
direction is aligned to a known polarized source operating from a balloon-borne
payload in the far field of ground telescopes. POLOCALC will fly in the Atacama
desert (Chile), within view of the Simons Observatory which includes the ACT
and Simons Array telescopes \cite{simons16}. The telescopes' detector
orientation is then measured by observing the calibration source signal. CMB
polarimeters will observe our artificial, well-characterized linear
polarization source in the far field of the fully integrated instruments
operating at cryogenic temperatures. The calibration sources will be visible at
high elevation angles, far from Earth signal contamination. Requirements of
such long distances, small zenith angles and high angular accuracy make this
measurement extremely challenging and never attempted before. Flying in the
Atacama region will maximize the impact of POLOCALC: any existing or future
project in the area, including ALMA and CLASS, will be able to use this novel
polarization calibrator which will enable measurements of the polarization
angle for each detector with respect to absolute sky coordinates.

\subsection{Methodology}
\begin{figure}[ht]
\begin{center}
\includegraphics[width=\textwidth]{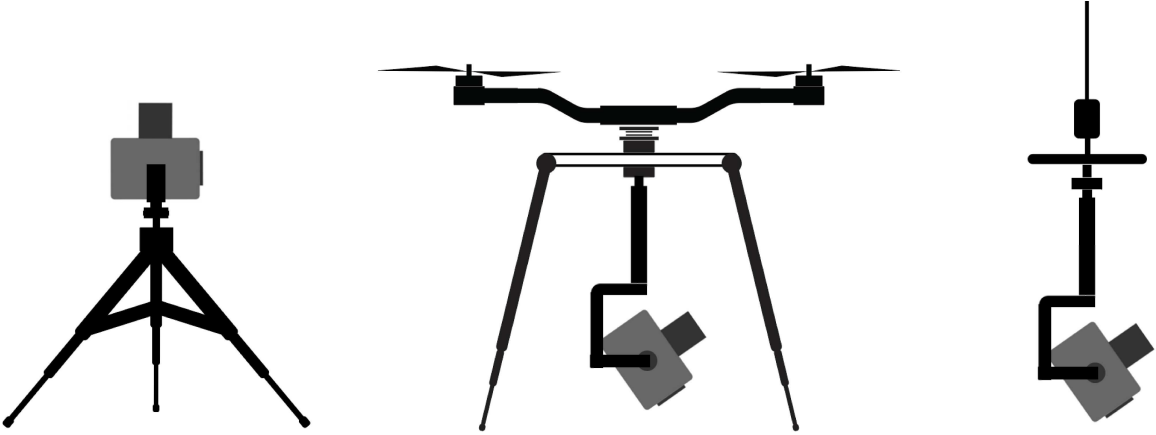} 
\end{center}
\caption{Three different configurations for POLOCALC (starting from the left):
1) from a ground tripod 2) from a flying drone and 3) from a high-altitude
balloon.}
\label{platforms:fig}
\end{figure}

POLOCALC will implement a novel calibration method to measure the polarization
angle and the telescope beam patterns making use of precisely characterized,
linearly polarized microwave sources. These artificial calibrators will match
the sensitive frequency bands of the CMB polarimeters and will operate from a
balloon-borne payload, illuminating the ground telescopes from far field
distances, appearing as distant sources for the polarization sensitive
detectors of the instruments. POLOCALC will be tested on the ground and on a
flying drone before being operated from a high-altitude balloon, as shown in
Fig. \ref{platforms:fig}. This will allow us to test the method with increasing
gradations of risk and performances, comparing far and near field results.

The calibration source must be placed at large distances and observed by the
telescopes at low zenith angles (high elevation angles). In the case of ACT,
POLARBEAR/Simons Array and Simons Observatory telescopes, the far field
distances range between 5 and 36 km for the sensitive bands centered between 40
and 150 GHz. For smaller telescopes, like CLASS, the far field distances are of
the order of 100 m. In a first test we will place the sources on nearby
mountains in sight of the telescopes, for example close to the TAO telescope on
the summit of Cerro Chajnantor. For small and medium size instruments, ground
sources will provide far field measurements. While this strategy already
crosses the accuracy thresholds of current methods, it comes with some
limitations. To observe a ground-based calibration source the telescopes must
point almost horizontally, so the polarimeters will receive a huge loading
background from the thermal emission of the Earth and the warm air mass. These
signals will saturate any typical CMB detector system which is optimized to
observe cold sky signals at temperatures lower than 10 K. Even if there are
methods to attenuate the loading on the detectors, they introduce additional
systematic errors affecting the calibration accuracy. Besides, there are
telescopes, like ACTPol, that cannot point at low elevation angles and require
at least 36 km of far field distance at 150GHz. A second test measurement will
overcome the ground-based limitations by mounting the calibration system on a
drone platform flying at ~100m above the telescopes. The drone is coupled with
a stabilization system that will reduce payload oscillations and will
constantly point the source towards the experiments. The source will have a
large beam, so pointing accuracy is not as important as attitude determination,
which will be provided by a separate system. For small aperture telescopes like
CLASS the drone already provides far-field calibration. For larger mirrors a
near-field measurement will also provide arcminute calibration when paired with
telescope optical models to simulate beam propagation
\cite{koopman16} and extrapolate the results for point-like sources.

The balloon-borne configuration will fully overcome all
existing limitations enabling both far field and high elevation (small zenith
angles) measurements for medium and large telescopes. POLOCALC will make use of
high-altitude balloons flying in sight of the experiments at around 30 km of
altitude.

\begin{figure}[ht]
\begin{center}
\includegraphics[width=0.7\textwidth]{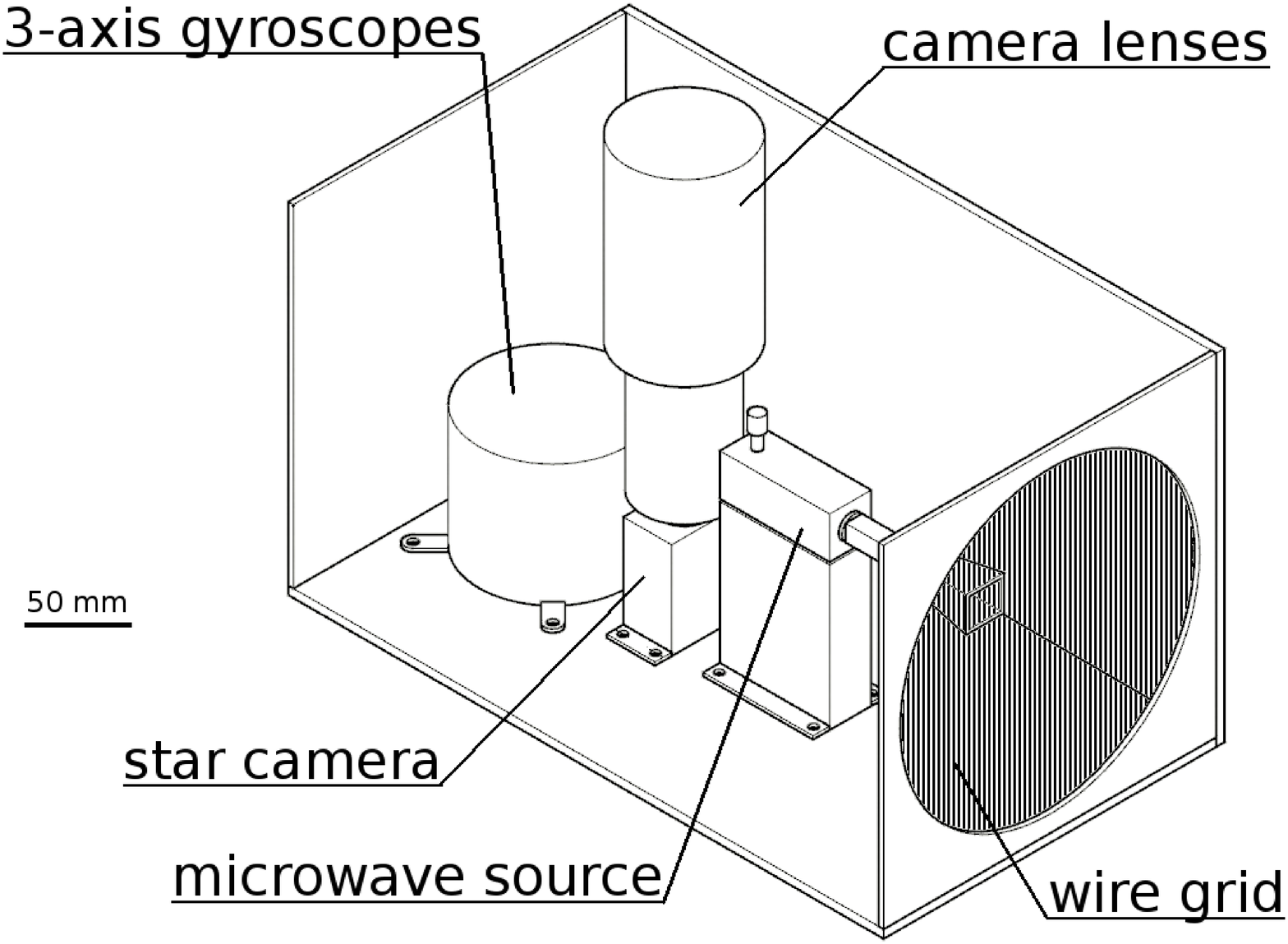} 
\end{center}
\caption{A schematic representation of the essential elements of POLOCALC. From
right to left: the linear polarization filter, the microwave source pointing
towards the telescopes, behind it the star camera looking at the sky (oriented
at 90$^{\circ}$. with respect to the source direction), and the 3-axis high-precision
gyroscopes.}
\label{polocalc_schematic:fig}
\end{figure}

POLOCALC will provide sources with polarization orientation registered in
absolute celestial coordinates with an accuracy between 0.01$^{\circ}$ and
0.001$^{\circ}$. We will measure the detectors' absolute orientation with
unprecedented accuracy and without cosmological model assumptions. The
systematics control will match the sensitivity levels expected for CMB-S4. The
absolute orientation of the polarized source will be provided with arcsecond
accuracy thanks to a state-of-the-art attitude control system (ACS) based on
star cameras, gyroscopes, and other orientation sensors like encoders,
clinometers, and magnetometers. A precisely machined frame will firmly hold
together the ACS star camera with the microwave source and filter, with the
respective pointing axis separated by 90$^{\circ}$, as shown in Fig.
\ref{polocalc_schematic:fig}. In this way, when the microwave source points
towards the telescope, the star camera will look at the sky. Our software will
recognize the stars' positions captured in the images and determine the
absolute orientation of the source. The pointing direction of the star camera
is aligned and parallel to the polarization direction of the microwave source
output with arcsecond accuracy. In other words, the camera pointing solution
also provides the orientation of the microwave source linearly polarized light.
This novel approach makes use of the high accuracy on the determination of the
two celestial angular coordinates of the star camera pointing solution to
determine the absolute orientation of the source polarized light with the same
exquisite arcsecond accuracy. While similar ground calibrations have already
been performed \cite{kaufman14a}, a sky referenced platform like POLOCALC has
never been used before. POLOCALC will make an unconventional use of
professional stabilization systems for drone video recording and hand-held
cameras to stabilize the pointing of the calibration source even on balloons.
POLOCALC's accuracy will be limited by the uncertainty of the star camera
solutions, the mechanical alignment precision of the linear polarization filter
direction of the source with the ACS, and the thermal stability and uniformity
of the payload. State-of-the-art ACS can typically provide an absolute
orientation determination with an accuracy between 0.01$^{\circ}$ and
0.001$^{\circ}$ \cite{gandilo14, chapman14}. The same accuracy range can be achieved for the assembly of
the camera and the wire-grid, making use of precise rotary stages, microscopes,
and accurate metrology systems. We will implement a thermal control providing
temperature stability between 2 K and 0.2 K, corresponding to an angular
misalignment on the order of 0.01$^{\circ}$ and 0.001$^{\circ}$ respectively.

It is important to highlight that we will transfer the calibration
obtained with POLOCALC among different detectors and telescopes by observing
the same sky signals, reducing the risk of calibration mismatch between
detectors. This is also useful both to compare the accuracy of different
instruments and to have the telescopes with ground screens limiting their view
to benefit from ground calibration. For example, POLARBEAR/Simons Array
telescopes can already make use of the ground measurements, and the results of
the calibration can be used for systematics control on the ACTPol instruments
via shared observation of celestial sources or CMB fields. Future telescopes
like the Simons Observatory will also be able to point at zero elevation.

POLOCALC's sky-referenced, balloon-borne polarization orientation calibrators will
overcome, one by one, the five main limitations of the current calibration
methods listed above, enabling:
\begin{arabiclist}[(5)]

\item Accurate calibration of the polarization angle for ground CMB
polarization experiments thanks to a precisely characterized distant artificial
source;

\item Calibration of fully integrated and cold receivers, avoiding compensation
for mechanical misalignments and differential thermal contractions between
laboratory and operating conditions;

\item Systematics control over thermal, optical or mechanical irregularities of
beam filling HWP or polarization filters. Calibration of the full optical
chain, including warm mirrors;

\item Near and far field measurements of the telescope beam patterns.
Calibrating the detectors' polarization angle and their polarized beams are two
different tasks, but for both of them a distant artificial source is
invaluable;

\item Calibration without assumptions on the primordial coupling of intensity
and polarization modes, as opposed to the self-calibration methods, thus
preventing induced biases on the cosmological parameters and enabling Cosmic
Birefringence measurements.

\end{arabiclist}

\subsection{Implementation}
\subsubsection{Sources}
POLOCALC will make use of multiple sources with frequency bands centered between
90 and 150GHz. We will utilize commercially available microwave oscillators,
horns and filters, but we will also develop and build custom emitters. The
sources will optimally match the brightness and frequency requirements of the
polarimeters. Frequency multipliers and filters will provide the desired output
frequency band.  Depending on the mirror size, source distance, detector noise
and frequency band, the final output power will be adjusted using the formulas
in \citet{lamarre86}. As an example, for sources at 90 and 150 GHz placed 5 km
away from the instrument, the required power is 500 pW and 30 pW respectively,
which means attenuation factors of 50/60 dB for a 50 mW Gunn oscillator. We will
couple the source to rectangular waveguides and horns to provide a linearly
polarized microwave light with -30dB of cross polarization level. A
polarization filter will be used to further clean or adjust the reference
signal. We will use precise photolithographed wire grid polarization filters.
The inside of the box containing the source will be covered with radio
absorbing material and we will also consider the use of mesh filters to
mitigate reflections between the filter and the horn. We will precisely align
and measure the polarization direction with arcsecond accuracy with respect to
the attitude control system camera. The frame holding together all the parts
requires a machining precision of 0.05 mm or better.
We will precisely characterize the microwave source in laboratory before deployment
The beam pattern of the source will also be precisely measured
using rotary stages, diode sensors and lock-in amplifiers. During the
operations in the field, the microwave source power will be modulated for
optimal signal recovery using a lock-in demodulation technique, while on the
telescope side we will also take advantage of the modulation systems such as a
rotating half-wave plate to change the polarization orientation.

\subsubsection{Attitude Control System}
The ACS is a custom composition of different subsystems with the scope of
measuring and controlling latitude, longitude, altitude and the 3-axis absolute
orientation of the polarized emitter. It is composed of a central processing
unit, optical sensors, attitude sensors, digital encoders and motors. POLOCALC
will use the same ACS on the ground, on the flying drone and on the
high-altitude balloon. We will utilize a state-of-the-art professional camera
stabilization system which make use of gimbals and motors to ensure high
rejections of the undesired trembling of the payload pointing. We will
stabilize the microwave source and the star camera with this system. While it
is designed for drones and hand held video recording, for the first time we
will adapt a camera stabilization system on a small balloon payload to point
steadily towards the telescope. Units like the FreeFly systems weight around 2
kg, have a battery life of several hours, and can be remote controlled or
programmed in target mode. We will protect and thermally isolate the
stabilization system's critical parts and test them in a environmental chamber
simulating the stratospheric environment. We will also develop and test an
independent pointing motor to stabilize the azimuth motion on the balloon
payload. The balloon moment of inertia is small, so reaction wheels will
provide enough moment of inertia and torque for azimuth movements. Due to the
short operating time at float it is unlikely that the reaction wheel will
saturate due to friction and balloon rotation, but we will consider the use of
a small azimuth motorized pivot to prevent this issue. Other custom solutions
to help to recover from reaction-wheel saturation can be explored if needed.
Attitude stability is a requirement for the star camera to take still pictures
of the sky, and for the microwave source in order to reduce the effect of
instrument time constants. POLOCALC will take advantage of improved and
miniaturized technology using a small, light, high sensitivity, high dynamic
range, low noise star camera coupled to professional lenses and a filter at 530
nm to enhance star signals over atmospheric noise. The static pointing of
POLOCALC allows frequent image acquisitions from the star camera. The
star-field images will be stored on a Solid State Drive which is light, low
consumption and does not require pressurization. Three-axis laser gyroscopes
will be integrated over time for fast and precise attitude reconstruction.
Further attitude sensors like clinometers, and magnetometers will contribute to
the alignment solution. We will use GPS time to synchronize the pictures and
the pointing solutions with the raw telescope data. During the flights
temperature gradients across the payload produce misalignments between the
filter and the camera boresight, and even uniform temperature variations of the
payload cause misalignments due to different materials of the aluminum holding
frame and of the dielectric substrate of the polarization filter. Variations of
2 K and 0.2 K correspond to misalignments of the order of 0.01$^{\circ}$ and
0.001$^{\circ}$ respectively. We will then utilize an active thermal control
providing the required temperature stability and uniformity within these
ranges. We will also insulate the vessel with foam and aluminized mylar layers.

\subsubsection{Flight Operations}
A high-altitude balloon with a diameter of $\sim$3 m and a weight of $\sim$ 2
kg can be hand-launched and take the payload to the stratosphere in around an
hour. A typical two hour flight will terminate at around 30 km from the
launching site. We will choose the launching site so that the ground
telescopes can observe the source at float at an elevation angle larger than
45$^{\circ}$. Line-of-sight and Iridium based telemetry systems will relay the
GPS position of the source to the ground. The telescope pointing systems will
utilize this information to track the source and perform beam maps.
Total payload weight will be limited to less than 10 kg by the drone and the
balloon restrictions. In January 2017, an aerial drone used to film the
telescopes at the Cerro Toco site showed feasibility for flying a stabilized
camera at around 100m above the experiments. We will adapt a similar drone to
our project. Stratospheric balloon flights will allow operations of the
calibrators from 30 km of altitude or more. POLOCALC balloons will fly above
the Atacama Desert in Chile, allowing visibility from all ground telescopes in
the area. We will select rugged hardware and protect all delicate parts for use
in the harsh conditions during deployment. In the stratosphere the operating
pressure is only a few mbar, so we will test all the parts in a vacuum chamber
in the laboratory. We will design the temperature control system to keep all
the parts in their working range. To allow optimal observations the balloon
will be launched during the night, but daytime flights are also possible in
case of restrictive wind conditions. Multiple pointing strategies will be
available thanks to the movements allowed by the ACS and the stabilization
system. Depending on the wind conditions, the payload will stay in sight of the
telescopes for few hours approximately. The lock-in technique and the high
signal to noise ratio from the source allow short integration times, so such a
flight duration is enough to calibrate the instruments and generate beam maps.
POLOCALC total consumption will be less than 25 W. For ground use either a lead
battery pack or lithium battery pack are suitable, while for the flight we will
make use of high capacity lithium battery packs, that can easily provide 150 -
200 W·h/kg. The balloon payload only requires a light, low consumption
telemetry to relay tracking data for the ground antennas' auto-pointing and to
allow access for all the telescopes that need to track the source. Drone
flights will also be used to test telescopes tracking systems. Airliners and
public risk are not of significant concern due to the operating area.

\section{Scientific Impact}
\begin{figure}[ht]
\begin{center}
\includegraphics[width=\textwidth]{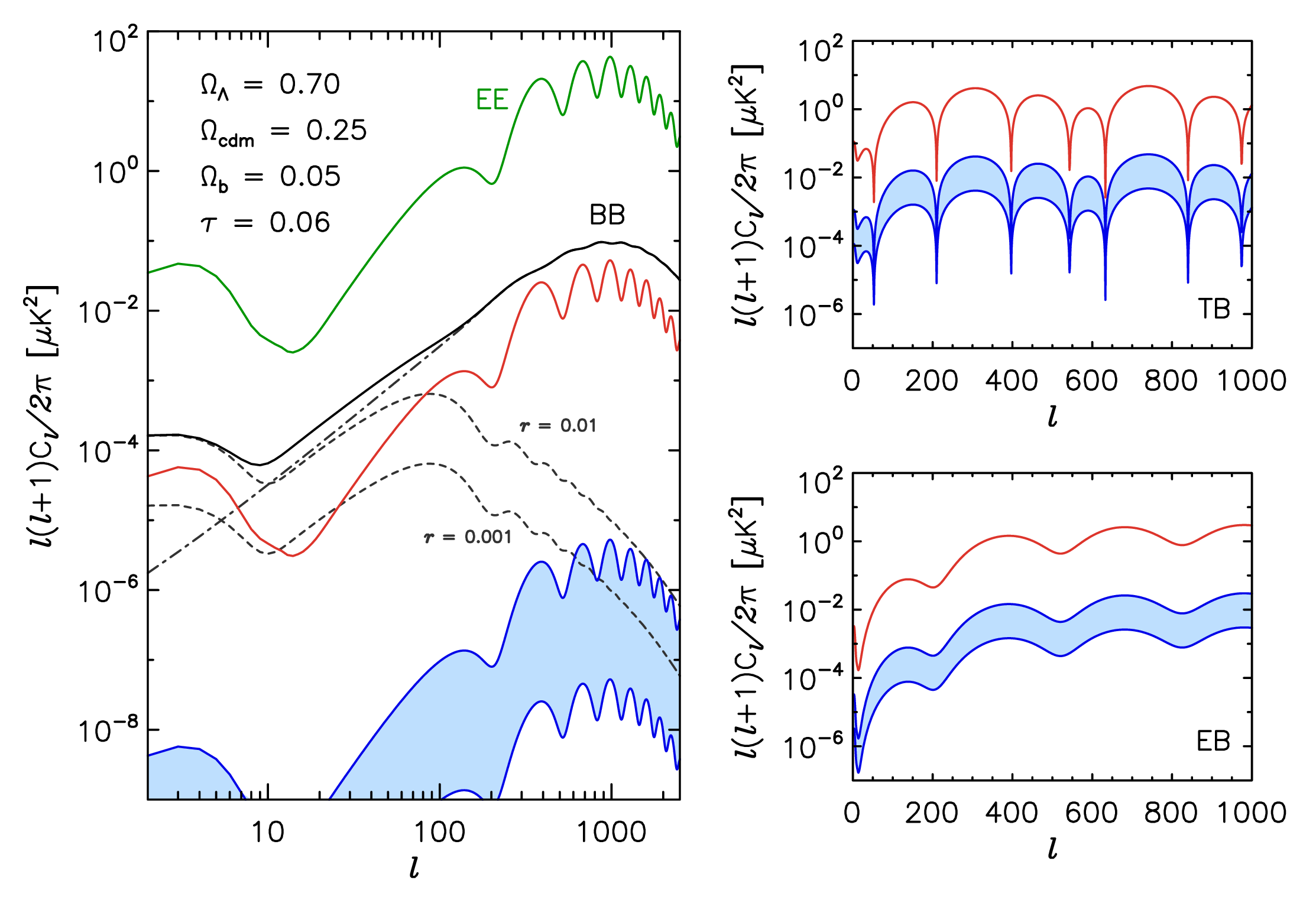} 
\end{center}
\caption{
Systematic errors in the BB, TB and EB power spectra produced by polarization
rotation. The red curves show the spurious signal produced by 1$^{\circ}$ of
rotation, which is approximately the current error level set by the systematic
uncertainty in available calibration sources. With an improved calibration
accuracy between 0.01$^{\circ}$ and 0.001$^{\circ}$, POLOCALC will allow any
rotation errors to be suppressed to the blue region in all plots. This
suppressed error level is negligible for current experiments targeting $r
\simeq 0.01$ and future experiments like CMB-S4 that will target $r \simeq
0.001$.  Primordial B-mode signals (dashed), the gravitational lensing B-mode
signal (dash-dot), and the EE signal (green) are plotted for reference in the
left panel.}
\label{spectra:fig}
\end{figure}

An anistropic rotation of the polarization angle produces a signal leak from
E-mode or temperature spectra to B-mode signals. It also produces a correlation
between these signals represented by the TB and EB cross spectra. The following
simulations show the effect of a common miscalibration angle, i.e. shared by
all detectors. In Fig. \ref{spectra:fig} we show the systematic errors in the
BB, TB and EB power spectra produced by polarization rotation. The red curves
show the spurious signal produced by 1$^{\circ}$ of rotation, which is
approximately the current error level set by the systematic uncertainty in
available calibration sources. With an improved calibration accuracy between
0.01$^{\circ}$ and 0.001$^{\circ}$, POLOCALC will allow any rotation errors to
be suppressed to the blue region in all plots. This suppressed error level is
negligible for current experiments targeting $r = 0.01$ and future experiments
like CMB-S4 that will target $r = 0.001$.  Primordial B-mode signals (dashed),
the gravitational lensing B-mode signal (dash-dot), and the EE signal (green)
are plotted for reference in the left panel. In the following sections we
present the expected scientific impact of POLOCALC, which can be preliminarily
evaluated by simulating CMB data with noise properties that match those
expected from current (Advanced ACTPol) and near future (CMB-S3) experimental
configurations. The posterior distributions are then obtained by performing a
Markov Chain Monte Carlo analysis employing the public COSMOMC package
\cite{lewis02} and an exact likelihood approach \cite{perotto06}.

\subsection{Inflationary Gravitational Waves}
\begin{figure}[ht]
\begin{center}
\includegraphics[width=\textwidth]{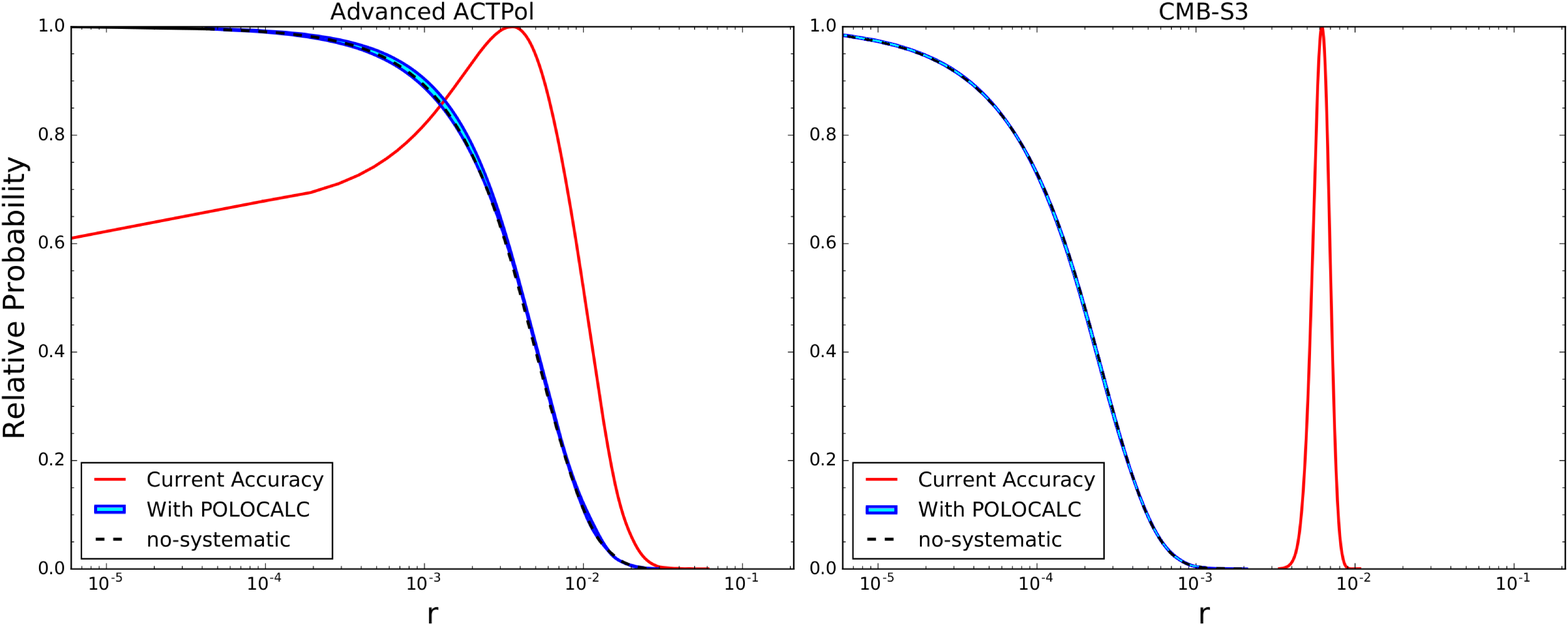} 
\end{center}
\caption{Assuming no gravitational waves, the red curve in these simulations
represents a false detection of $r$ caused by the polarization angle
miscalibration of 1$^{\circ}$. With a calibration accuracy between 0.01$^{\circ}$ and 0.001$^{\circ}$
represented by the blue region, POLOCALC recovers the fiducial value of r=0
(the black dashed curve). The uncertainty on the value of $r$ will be then limited by
the sensitivity of the experiment. The false bias already starts to emerge
above the statistical uncertainty for the ACTPol sensitivity, while for CMB-S3
it dramatically affects the measurement of the tensor-to- scalar ratio. For
CMB-4 the gap is going to increment.}
\label{plot_r:fig}
\end{figure}

A miscalibration of 0.5$^{\circ}$ in the polarization orientation translates
into a spurious B-mode signal corresponding to a tensor-to-scalar ratio of $r
\simeq 0.01$ \cite{abitbol16}, affecting the sensitivity range of existing and
planned experiments. POLOCALC will calibrate the polarization angle with
arcsecond accuracy. As a consequence, the uncertainty on the value of $r$ will be
limited by the sensitivity of the experiment. A simulated result is shown in
Fig. \ref{plot_r:fig}, for an ACT-like and a CMB-S3 experiment, where the
red curve represents a false bias signal introduced by a miscalibration of
1$^{\circ}$ (i.e., the current accuracy) in the orientation of the detectors.
The bias starts to emerge above the statistical uncertainty for ACTPol
sensitivity (the image on the left). For CMB-S3 (the image on the right), the
new generation of ground experiments, the bias is well above the sensitivity
and it dramatically affects the measurement of r. If the same experiment
benefits from the POLOCALC calibration, gaining an accuracy between
0.01$^{\circ}$ and 0.001$^{\circ}$, the bias can be recovered as represented by
the blue region. For CMB-S4 the gap between accuracy and sensitivity will
increase, so the effect on cosmological parameter estimations will be even more
important. 

\subsection{Lensing and foregrounds}
\begin{figure}[ht]
\begin{center}
\includegraphics[width=\textwidth]{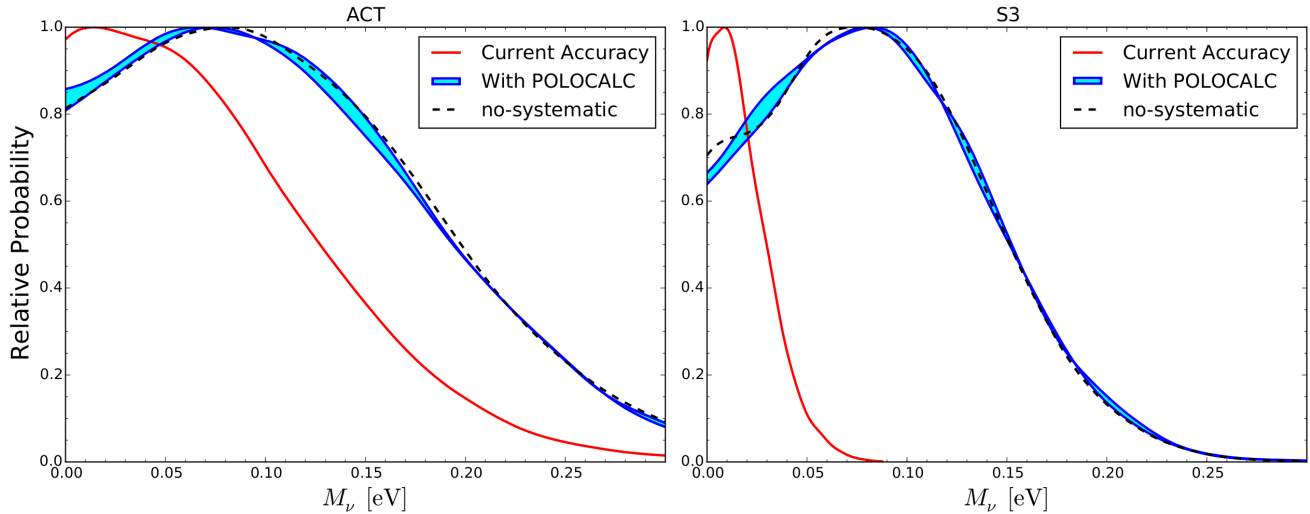} 
\end{center}
\caption{The red curve in these simulations represents a wrong estimation of
the sum of neutrino masses (in this model assumed as 0.06 eV) caused by a
polarization angle miscalibration of 1$^{\circ}$ for AdvACTPol-like and CMB-S3
experiments. With a calibration accuracy between 0.01$^{\circ}$ and
0.001$^{\circ}$ represented by the blue region, POLOCALC recovers the fiducial
value (the black dashed curve). Such a gap is going to be more relevant for CMB-S3 and
CMB-S4.}
\label{plot_mnu:fig}
\end{figure}

While CMB photons leave the Last Scattering Surface shortly after the Big Bang,
matter distribution introduces gravitational lensing at later times. However,
on the one hand E-modes could be converted into B-modes through a polarization
plane rotation before lensing happens. On the other hand non-vanishing TB
and EB spectra from local Galactic foregrounds can be present in the
observations. In these cases lensing measurements will suffer from an
uncontrolled bias. This would affect both the lensing potential reconstruction
and the delensing efficiency. POLOCALC calibration does not require assumptions
on the polarization rotation history or on the correlations between the
spectra. Therefore, POLOCALC will enable gravitational lensing potential
measurements limited by statistical uncertainties. The sum of neutrino masses
is one of the most relevant parameters encoded in CMB lensing signals. In Fig.
\ref{plot_mnu:fig}, the black dashed curve represents the probability distribution of
the sum of the neutrino masses as it would be constrained by an experiment as
sensitive as ACT, if systematic effects (such as the error in the
orientation of the detectors) are perfectly under control. The red curve
clearly shows the bias induced by a miscalibration of 1$^{\circ}$ (i.e., the
current accuracy) in the orientation of the detectors. If the same experiment
benefits from the POLOCALC calibration, gaining an accuracy between 0.01$^{\circ}$ and
0.001$^{\circ}$, the bias can be recovered as represented by the blue region. 

\subsection{Cosmic Birefringence}
Detecting a Lorentz- and parity-violating process like CB would be a paradigm
shift in physics. The outcome of a similar discovery would be revolutionary,
and its consequences would deeply affect our understanding of fundamental
processes and the description of our Universe. Since a polarization angle
miscalibration is indistinguishable from CB, self-calibration methods come at
the price of losing sensitivity to Cosmic Birefringence. POLOCALC calibration
will enable measurements of the CB rotation angle, $\alpha_{CB}$, providing an
independent, experimental calibration method with an unprecedented accuracy on
the absolute polarization orientation between 0.01$^{\circ}$ and
0.001$^{\circ}$, reaching the limit of statistical uncertainties required by
CMB-S4.

\subsection{Primordial Magnetic Fields}
While CB is a departure from the Standard Model, Faraday Rotation is a
well-known physical effect that may be used to measure the intensity of
Primordial Magnetic Fields in the early Universe and select cosmological
models. PMFs could also explain the presence of magnetic fields and their
evolution in the Universe. The FR angle affecting the polarization plane of CMB
photons is directly proportional to the magnetic field intensity. POLOCALC will
improve the accuracy on the polarization orientation from 1$^{\circ}$ to better than
0.01$^{\circ}$, therefore scaling by two orders of magnitude the sensitivity to
Primordial Magnetic Fields through FR with respect to current possibilities.

\subsection{Additional advantages}
From an experimental point of view, the importance of using a fully
characterized source to illuminate the telescopes goes even beyond the scope of
the polarization angle calibration and beam functions characterization.  CMB
experiments make use of technologies with frequency dependent performances
within the broad sensitivity bands of the polarimeters. For example,
technologies like achromatic half-wave plates
\cite{pisano14} and sinuous antennas have
frequency dependent performances. While the frequency spectrum of a celestial
source is not precise enough to recover these in-band features, both POLOCALC
narrowband and broadband sources have the advantage of being fully characterized
in laboratory, before deployment. Therefore, they can be used to study finer
in-band instrumental features during the observational campaigns. POLOCALC
clean and narrow band signals can be used to compare the polarimeter transfer
functions during operations with the lab-measured ones. POLOCALC can also be
used to measure the modulation efficiency and polarimeter axis angle curves for
an achromatic half-wave plate polarimeter, or the polarization angle dependency
of a broadband sinuous antenna. Another important advantage of POLOCALC is to
provide a calibration source to inter-calibrate each polarization sensitive
detector of the same instrument, and even across different telescopes.

\subsection{Advantages over a CubeSat}
While a low Earth orbit small satellite like a CubeSat could also provide a
distant calibration source \cite{johnson15}, an orbital mission has the following disadvantages
with respect to the POLOCALC balloon-borne calibrators:

\begin{enumerate}
\item Assuming a polar orbit with an altitude of 500 km, a CubeSat-based
instrument would be visible from the telescopes in the Atacama Desert for only
about 2 minutes in a given orbit and for only a few times each week.  The
satellite would need to be tracked by the telescopes while it crosses the sky,
but the angular speed of the satellite can at times exceed 0.6 $^{\circ}$/s, which is a
common upper limit for telescope mounts. Therefore, making beam maps will be
challenging.
\item CubeSats pose much stronger limitations in size, weight and power. These
constraints limit the calibration source technologies that can be used.  Also,
the broadcast frequency bands are restricted by international regulations, so
the available frequencies may not optimally match the spectral bands in the
polarimetric receivers.
\item Finally, a satellite is a high risk enterprise, it can be more expensive,
and it would not be available immediately.
\end{enumerate}

\section{Conclusion}
POLOCALC has the potential to become a rung in the calibration ladder for
existing or future CMB experiments observing our novel polarization calibrator.
This novel method will enable measurements of the polarization angle for each
detector with respect to absolute sky coordinates with unprecedented accuracy.
This project will produce the first independently calibrated measurement of the
polarization angles of the CMB light and its contaminants allowing Cosmic
Microwave Background polarization experiments to fully mine the cosmic sky.

\end{document}